\documentclass[prc, twocolumn]{revtex4-1}
\usepackage{amsmath}
\usepackage{graphicx}
\usepackage{xcolor}
\def\lsim{\mathrel{\raise3pt\hbox to 8pt{\raise -6pt\hbox{$\sim$}\hss{$<$}}}}

\begin{document}
\bibliographystyle{apsrev}
\hfill \fbox{\parbox[t]{1.25in}{LA-UR-14-29105}}
\title{Reaction-in-Flight Neutrons as a Test of  Stopping Power in Degenerate Plasmas }

\author{A.C. Hayes, Gerard Jungman, A.E. Schulz, M. Boswell, M.M. Fowler, G. Grim, A. Klein, R.S. Rundberg, J.B. Wilhelmy, D. Wilson
}
\affiliation{Los Alamos National Laboratory, Los Alamos, NM, USA  87545}
\author{C. Cerjan, D. Schneider, A. Tonchev, C. Yeamans
}
\affiliation{Lawrence Livermore National Laboratory, Livermore, CA, 94551}
\begin{abstract}
We present the first measurements of reaction-in-flight (RIF) neutrons in an inertial confinement fusion system.
The experiments were carried out at the National Ignition Facility, using both Low Foot and High Foot drives and
cryogenic plastic capsules. In both cases, the high-energy RIF ($E_n>$ 15 MeV) component of the neutron spectrum 
was found to be about $10^{-4}$ of the total. The majority of the RIF neutrons were produced in the 
dense cold fuel surrounding the burning hotspot of the capsule and the data are consistent with a compressed cold fuel that
is moderately to strongly coupled $(\Gamma\sim$0.6) and electron degenerate $(\theta_\mathrm{Fermi}/\theta_e\sim$4). The production of RIF neutrons is controlled by
the  stopping power in the plasma. Thus, the current RIF measurements provide a unique test of stopping power models in an experimentally unexplored plasma regime.
We find that the measured RIF data strongly constrain stopping models in warm dense plasma conditions and some models are ruled out by our analysis of these experiments.
\end{abstract}
\maketitle
\section{Introduction}
At the National Ignition Facility (NIF)\cite{NIF}, high yield inertial confinement fusion (ICF) plasmas are studied using
cryogenically cooled capsules containing an equimolar mixture of deuterium and tritium (DT).
The cryogenic capsule designs involve three distinct material layers: an outer hydrocarbon ablator layer, 
a thick layer of solid DT ice, and a central sphere of DT gas. 
The capsule is compressed to sufficient temperatures and densities that burn is initiated in the central gas region with a design that promotes
the propagation of a thermal wave into the surrounding dense cold DT fuel. 
Though burn propagation into the cold fuel has yet to be achieved, the assembled 
configuration of the compressed capsule is suitable for the production of rare nuclear reactions
with measurable yields.
One such process, the production of reaction-in-flight or tertiary neutrons, is the central focus of the present work.
We examined implications of the measured yield of these high-energy neutrons within the context of 
different theoretical stopping power models. 

Reaction-in-Flight (RIF) neutrons require three successive reactions.
First, a primary 14.1 MeV DT neutron is produced; second, this high energy neutron then undergoes elastic scattering with 
deuterium or tritium ions in the plasma, energetically up-scattering these ions to a range of energies from
0 to more than 10 MeV. In the third step, the energetic
knock-on ion undergoes a DT reaction with a thermal ion in the plasma,
producing a continuous spectrum of RIF neutrons in the energy range 9.2-30 MeV. 
 The process of RIF production is summarized in Figure 1.
Knock-on ions lose energy as they traverse the plasma, which directly affects the number and energy spectrum of
the produced RIF neutrons.
Thus, this sensitivity can be used to extract information about plasma stopping powers. As detailed below, in the case of 
cryogenic NIF capsules, RIFs probe stopping under previously unexplored plasma conditions. 
\begin{figure}
\includegraphics[width=2.5in, bb= 0 0 792 612]{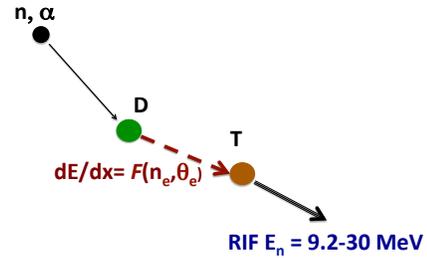}
\caption{RIF production involves three consecutive steps. 
First, a primary neutron and alpha particle are produced in a DT fusion reaction. Either of these can then 
knock a D or T ion up to MeV energies through elastic scattering. 
The knock-on ion then undergoes a secondary DT fusion reaction producing a RIF neutron.  
The specific RIF diagnostic neutrons in the present work are those measured by the $^{169}$Tm$(n,3n)$ reaction, 
for which $E_n>$15 MeV. This Tm diagnostic is dominated by the high-energy neutron-induced RIF since
the alpha-induced RIFs only contribute about 10\% of the total thulium signal.}    
\end{figure}

\section{ RIF production in cryogenic NIF capsules}
The RIF production depends directly on the fluence of knock-on ions in the plasma. 
Under plasma conditions in which they can escape the capsule, knock-on ions have been successfully measured and
characterized, including their partial down-scattering \cite{MFE}.
The magnitude of the initial  
knock-on fluence $Q_0$ is determined  by the  plasma burn conditions, while the spectral shape, $q_0(E_0$), is determined simply by  the kinematics of
the elastic collisions between the 14 MeV neutrons and the plasma ions.
The magnitude of the knock-on fluence is controlled by the 14 MeV neutron fluence, $\phi_n$, the number density of DT 
ions in the plasma, $\hat{n}_{DT}$, 
and the knock-on cross section $\sigma_{KO}$, i.e., $Q_0=\phi_n\hat{n}_{DT}\sigma_{KO}$.
Thus, a capsule with a high 14 MeV fluence and a high DT density is needed to ensure a high knock-on fluence.
For this purpose, the cryogenic designs are ideal: they provide a strong neutron source (currently up to $10^{16}$ total neutron number) 
from the burning hotspot, and a high number density ($10^{25}$-$10^{26}$ cm$^{-3}$) of DT ions in the cold fuel.  
From these simple considerations, we expect the RIF production to take place predominantly in the cold fuel, and 
this expectation is realized in the detailed simulations discussed below.
 
Transport of the knock-on ions in the plasma results in a change from the initial shape $q_0(E_0)$ and
magnitude $Q_0$ of the spectrum, a change that is determined by the form of the stopping power.
If the stopping length of the knock-on ions is short compared to the spatial variations in the temperature
and density within the plasma, the differential knock-on fluence can be related \cite{stopping-jungman}
to the energy loss, $dE/dx$, and the initial knock-on fluence by
\begin{equation}
\frac{d\psi_{ko}}{dE_{ko}}(E_f) = \frac{Q_0}{|dE/dx(E_f)|} \int^{E_{0max}}_{E_f}\;dE_0\; q_0(E_0).
\label{bigeqn}
\end{equation}
From equation \ref{bigeqn}, it follows that determination of the knock-on fluence  in the plasma is
inversely related to the stopping power.
In Figure 2 we show the shape of the knock-on fluence for deuterons in a plasma with constant density and temperature, 
$n_e=10^{26} $cm$^{-3}$ and $\theta_e=0.2$ keV, using three different stopping power models and eq. \ref{bigeqn}. 
The upper curve uses the Maynard-Deutsch \cite{MD} stopping power as parameterized by Zimmerman \cite{zim}, while the dashed
curve uses the same model but neglects electron degeneracy. Electron degeneracy lowers the magnitude of the
stopping, increasing the knock-on fluence. We discuss the effects of degeneracy in more detail in the next section.
Finally, the lowest curve in Figure 2 shows the effect of using a purely classical stopping model \cite{paul}; as can be seen, quantum effects (including degeneracy) increase the knock-on fluence by more than a factor of two over classical predictions.
\begin{figure}
\includegraphics[width=3in, bb= 0 0 792 612]{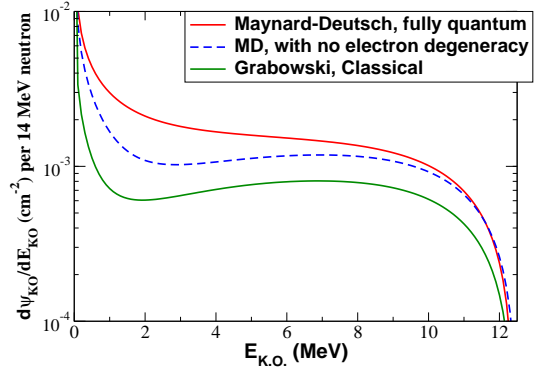}
\caption{The knock-on fluence for deuterons in a plasma with constant density and temperature, $n_e=10^{26}$ cm$^{-3}$ and $\theta_e=0.2$i keV, using different stopping power models and eq. \ref{bigeqn}.
The uppermost curve uses a fully quantum stopping model \cite{MD}, while the dashed
curve uses the same model but neglects electron degeneracy. 
The lowest curve is the prediction from a purely classical stopping model \cite{paul}.
Quantum effects (including degeneracy) increase the knock-on fluence by more than a factor of two over classical predictions.}
\end{figure}

 We note that, in general, it is not necessary to invoke the assumption that the stopping lengths are short, and our full simulations make no such assumption. 
Nonetheless, eq. \ref{bigeqn} is very useful in illuminating the relationship between the stopping power and knock-on fluence.
The knock-on fluence, and hence the stopping, can be determined by measuring the yield of nuclear
reactions involving knock-on particles \cite{beta-mix}. In the present work we accomplish this goal by 
directly measuring the RIF neutrons.

The RIF production per volume is determined by the integral of the knock-on fluence over the DT cross section, weighted by the density of DT ions \cite{hayes-rif, density},
\begin{equation}
  \frac{d\Gamma_{RIF}}{dE_{RIF}dV} =
    n_{DT}\int_0^{E_{_{ko}}^{max}} \frac{d\psi_{ko}}{dE_{ko}}\sigma_{DT}\frac{dF}{dE_{RIF}}dE_{ko}
\end{equation}
The factor \(dF/dE_{RIF}\) is the kinematic factor needed to ensure that a given knock-on energy produces
 a RIF neutron within the measured energy window, which for the present experiments is the window
 $E_{RIF}>$15 MeV.
Again, because the density of DT ions is highest in the cold fuel, the majority of RIFs are produced in this
region of a cryogenic capsule.

The  RIF production at peak burn, predicted from  1-D \cite{cpt-burn} and 3-D \cite{hydra-burn} simulations, originates in the shocked cold fuel with fewer than 10\%  produced in the hot spot.
Figure 3 presents plots of the simulated spatial dependence of the energy deposition of the knock-on deuterons from a 3-D HYDRA \cite{hydra-burn} simulation. 
The majority of deuterons deposit their energy in the cold fuel. For comparison, we show the position of alpha-particle energy deposition, which is entirely contained in the hotspot of the capsule.
While both experiment and simulations show capsule-to-capsule variations in the neutron yield and the density and temperature of the cold fuel, 
the situation depicted in Figure 3 is representative of the conditions in cryogenic capsules.
\begin{figure}
\includegraphics [width=3in, bb= 0 0 720 540]{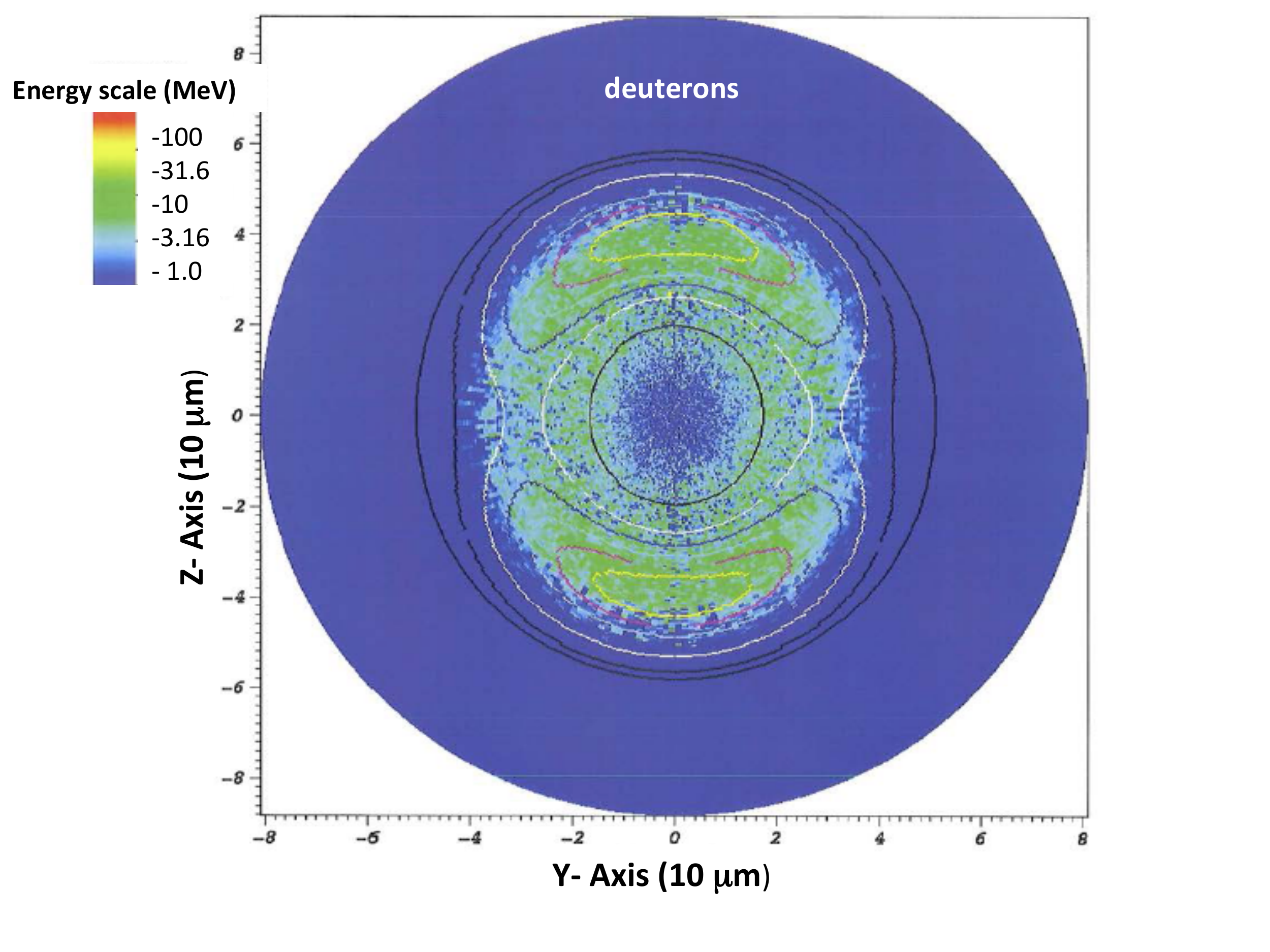}
\includegraphics [width=3in, bb = 0 0 720 540]{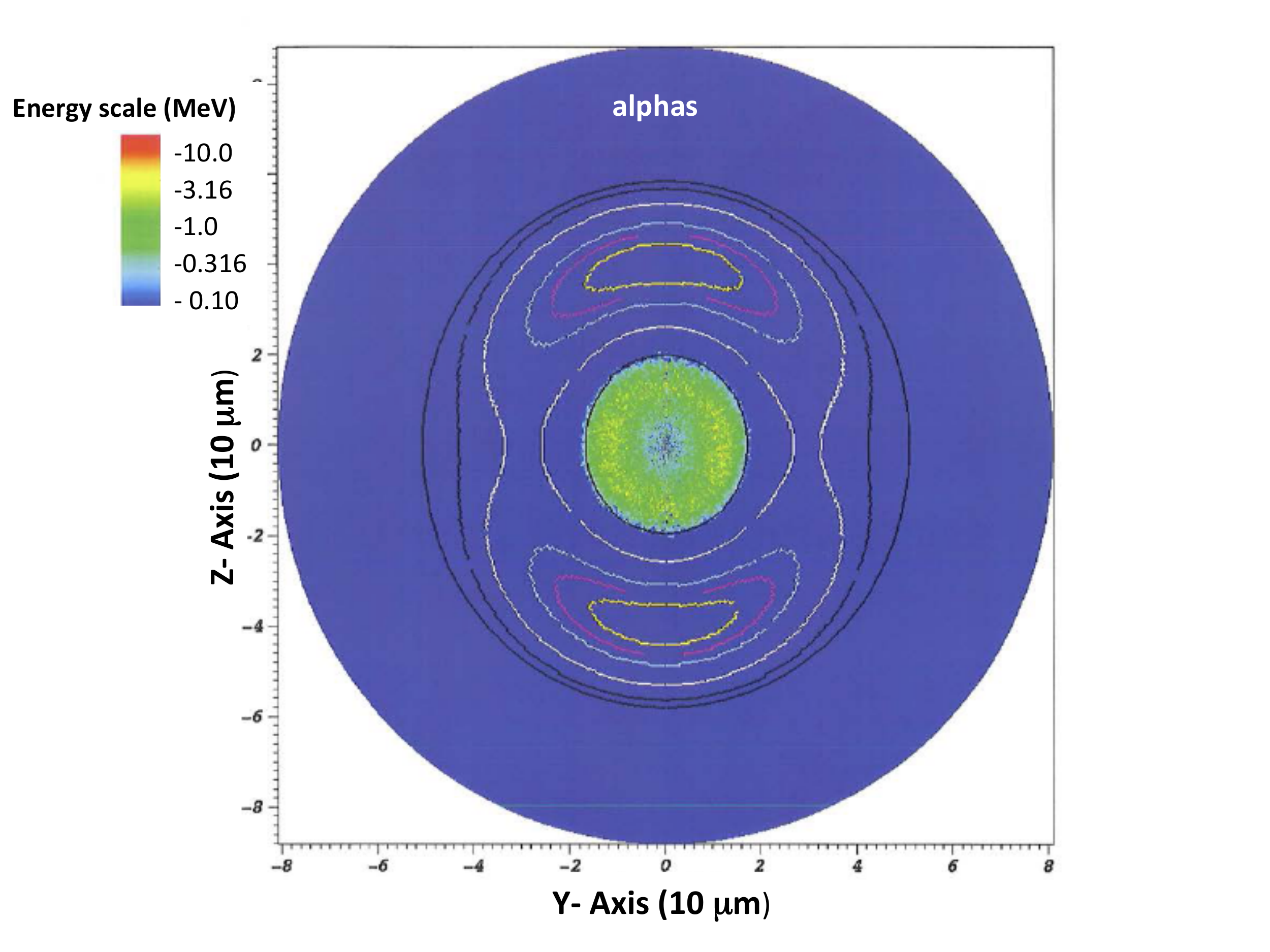}
\caption{The predicted spatial dependence of the energy deposition of the knock-on deuterons and alphas 
from a 3-D HYDRA \cite{hydra-burn} simulation with superimposed density contours for a typical cryogenic capsule implosion relevant to the present
RIF measurements. The knock-on deuterons deposit their energy predominantly in the cold fuel. In sharp
contrast, the alpha particles deposit all of their energy at the edge of the hotspot.
}
\end{figure}
\section{The plasma coupling and stopping Models}
At peak capsule compression, the outer dense DT fuel is at a temperature of a fraction of a keV and a density that is approaching  10$^{26}$ cm$^{-3}$. 
At these densities and temperatures, some standard stopping models, such as the Brown, Preston and Singleton (BPS) model \cite{BPS}, 
become inaccurate because such models are designed for and restricted to weakly-coupled plasmas. 
The plasma coupling parameter is defined as the ratio of the potential energy to the plasma temperature,
$\Gamma= Ze^2/\theta R$, where $Z$ is the charge of the moving ion,  $\theta$ is the plasma temperature, and $R$ is a radius that roughly describes the distances between charges in
the plasma. 
Different choices for $R$ are found in the published literature, with the Wigner or Debye radius being the most common choices. 

Loosely speaking, low densities and high temperatures correspond to weakly coupled plasmas in
which long-distance collective effects dominate, while high density, low temperature plasmas tend to be strongly
coupled. 
For small values of $\Gamma$ ($\Gamma<0.1$) and non-degenerate plasmas, there are several stopping models that are known to be accurate, see for example \cite{BPS, Zwick, MD}. 
Though the theoretical framework used is quite different in each of these treatments, the derived stopping powers can be shown \cite{stopping-test} to be numerically identical under weakly-coupled conditions.
As the coupling $\Gamma\rightarrow 1$, though, the plasma becomes strongly coupled. 
An additional important characteristic of the plasma that affects the form of the energy loss is the degree of electron degeneracy which
is determined by the ratio of the electron temperature $\theta_e$ to the Fermi energy $\theta_F$.
For a degenerate plasma, the temperature determining the stopping power is an effective temperature,
$\theta_\mathrm{eff}=\frac{3}{5}\theta_F F(\theta_e)$.  The function $F(\theta_e)$ is obtained by solving for the chemical potential of the system 
and has the property that in the zero temperature limit,
$\theta_\mathrm{eff}\rightarrow \frac{3}{5}\theta_F$, while in the high temperature limit $\theta_\mathrm{eff}\rightarrow\theta_e$. 
Under moderately to strongly coupled conditions, $\Gamma>0.3$, published models of plasma stopping powers disagree, 
especially since not all the models include degeneracy effects. To date, there have been no
experimental tests of plasma stopping powers for non-weakly coupled and degenerate plasmas. However,
 the cold fuel of cryogenic NIF capsules do provide a system with such conditions, hence the RIF measurements
probe stopping powers in a previously unexplored plasma regime. 

Figure 4 contains a plot of the ratio of the Fermi temperature to the actual temperature at peak burn time 
as function of the capsule radius, as predicted in a 1-D HYDRA simulation for a standard High Foot\cite{highfoot} cryogenic capsule implosion.
As can be seen, the cold fuel is degenerate, with the electron temperature being a factor of 3-5 below the Fermi temperature. 
The cold fuel is divided into an inner shocked and outer unshocked region, the inner shocked region being higher in density and temperature.

Figure 5 displays the effective temperature and  the ratio $\rho/(\theta_\mathrm{eff}^{3/2})$. This  ratio controls the magnitude
of the stopping power, which in the case of stopping by electrons, can be parameterized as
\begin{equation}
\frac{dE}{dX}=-\frac{4\pi Z_p^2 e^4}{m_e v_p^2}\left(\frac{n_e}{\theta_e}\right)G(y) ln(\Lambda_e)\;,
\end{equation} 
where $v_p$ and $Z_p$ are the velocity and charge of the projectile, $n_e(\equiv\rho)$ is the electron density, and $G(y=\frac{m_e E_p}{m\theta_e})$ 
is a function controlling the shape of the stopping\cite{MD,LiP}. 
For small $y$, $G(y)\sim y^{1/2}$ and the magnitude of the stopping power is set by the ratio
$n_e/(\theta_\mathrm{eff}^{3/2})$.
In a fully degenerate system, the effective temperature is fully determined by the density, and the ratio
$n_e/(\theta_\mathrm{eff}^{3/2})$ is a fixed physical constant.
For this reason,  the stopping power is more constrained in the cold fuel, and a cold fuel 
 temperature measurement is not necessary. This is in strong contrast to a weakly coupled system.
\begin{figure}
\vspace*{1cm}
\includegraphics[width=3in, bb =0 0 792 612]{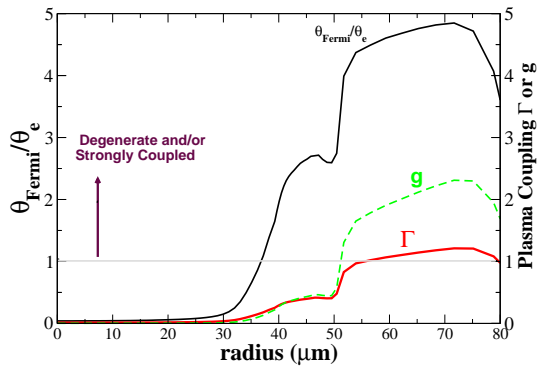}
\caption{The ratio of the Fermi temperature to the electron temperature (black curve) as a function of the capsule radius at peak burn time from a 1-D HYDRA simulation. 
In the cold fuel (CF) ($r\sim 40-80 \mu m$), the plasma is electron  degenerate. 
Also shown  are two standard definitions of the plasma coupling, $\Gamma=\frac{Ze^2}{\theta R_{W}}$(red curve) and $g=\frac{Ze^2}{\theta R_{D}}$ (green curve), 
where $R_W$ and $R_D$ are the Wigner and Debye radii respectively. The cold fuel ranges from moderately coupled to strongly coupled.
At peak burn the simulations predict that the CF has two distinct regions with different degrees of degeneracy and coupling, 
the region $r\sim40-50 \mu m$ within the shocked CF and the region $r\sim50-80 \mu m$ upstream of the shock. 
The capsule hydrocarbon ablator material starts at and extends beyond a radius of $80 \mu$m. }
\end{figure}
\begin{figure}
\vspace*{1cm}
\includegraphics[width=3in, bb = 0 0 792 612]{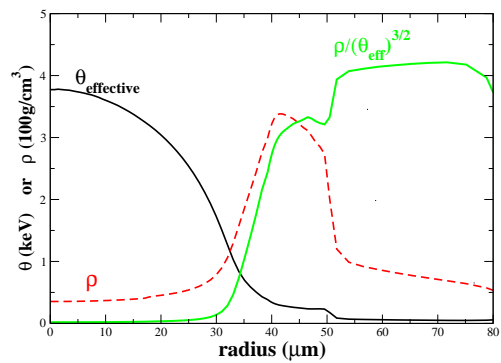}
\caption{ The effective temperature and the density at peak burn time, as a function of the DT fuel radius, from a 1-D HYDRA simulation.
 The ratio $n_e/(\theta_\mathrm{eff}^{3/2})$ is approximately constant in each region of the CF. 
It is this combination of density and effective temperature that determines the stopping power at low knock-on energies.
The electron degeneracy and the corresponding approximately constant value of $n_e/(\theta_\mathrm{eff}^{3/2})$
in the CF  greatly reduce the uncertainty with which we can extract $dE/dX$ in the absence of a CF temperature measurement.} 
\end{figure}
\section{The Experiments}
The signal from  RIFs is generally expected \cite{Hayes-pop} to be small $(\sim10^{-4})$ relative to the primary 14 MeV signal.
Thus, experimental detection requires measuring the neutron spectrum above the thermally broadened primary 14 MeV peak.
In the experiments reported here, RIF detection was achieved at NIF by measuring neutrons with energies above 15 MeV via neutron activation of thulium foils.
The $(n,3n)$ reaction on natural thulium ($^{169}$Tm) is a threshold reaction, requiring
$E_n>15$ MeV and producing $^{167}$Tm. The experiments were designed to search for the presence of $^{167}$Tm, which decays with a half-life $\tau_{1/2}=9.25$ days by electron capture to
$^{167}$Er, with the emission of a 207.79 keV $\gamma$-ray 41\% of the time.
 Thulium foils were placed in the  Solid Radiochemistry Collector (SRC) holders that are attached to a  Diagnostic
Instrument Manipulator (DIM) in the NIF target chamber \cite{Dawn}.
 The SRC holders are  50 cm from the ICF target, and two Tm foils were used, one in a holder on a DIM in the equatorial direction  and one in the polar direction.
Thulium foils of thickness 2.0 mm or 0.5 mm were used in the experiments depending on the shot.
The quantitative measurement of $^{168}$Tm provided a consistency check of the primary 14 MeV neutron production 
that was determined independently by several direct time-of-flight measurements, as well as other activation techniques.

The primary experimental challenge for the $^{167}$Tm  measurements was the huge background of $\gamma$-rays produced in the activation of the foils by
the much higher fluence of primary 14 MeV neutrons, in particular those from $^{168}$Tm $(\tau_{1/2}$=93.1 days).
Thus,  suppression of background gamma-rays was crucial in the experiments.
To achieve this suppression, the activated foils were shipped to Los Alamos (LANL), where two clover detectors were 
deployed to assay the $^{168}$Tm and $^{167}$Tm activity in the foils.
The LANL clover system consists of two high efficiency clover Ge detectors.
Each clover consists of 4 segmented high purity Ge crystals, surrounded by an active NaI Compton suppressor.
The $^{167}$Tm decays to a 208 keV isomeric level in $^{167}$Er that has a half-life of 2.28 sec.  
This level decays 100\% of the time to the $^{167}$Er ground state.
 Since there is nothing in coincidence with this transition, this decay sequence is ideal for identification 
by our segmented clover detector.  
In contrast, the major background from the decay of $^{168}$Tm has a highly complex decay sequence 
that has a high probability of interacting with multi-elements in the clover array.    
These coincident events can be readily removed from the event stream, resulting in a greatly improved $^{167}$Tm signal-to-noise ratio.
Details of the setup are presented in \cite{gary}.

The first positive RIF signals were seen in March of  2013 for the NIF shot N130331.
These signals were obtained from a cryogenic implosion in which the protocol was a design from the National Ignition campaign (NIC), sometimes referred to as a Low Foot design \cite{NIF}.
The measured average $^{167}$Tm/$^{168}$Tm ratio for the two 2 mm foils was (2.0$\pm$0.74)$\times$10$^{-5}$, 
which corresponds to a RIF($E_n>$ 15 MeV)/total neutron ratio of 10$^{-4}$. 
The measured $^{167}$Tm/$^{168}$Tm atom ratio is the integral of the RIF/total spectrum 
against the respective Tm isotopic production cross sections. 
Since these cross sections are adequately known, the integral primary RIF spectrum can be deduced. 
During 2013, three additional RIF  measurements were performed 
using 2 mm Tm foils, and the results 
were in the 1-2$\times10^{-5}$ range, with an uncertainty of about 35\%.
These three additional shots were High Foot design \cite{highfoot} capsules.
The results for the four 2 mm foils are summarized in Figure 6.
\begin{figure}
\includegraphics[width=3in, bb= 0 0 792 612]{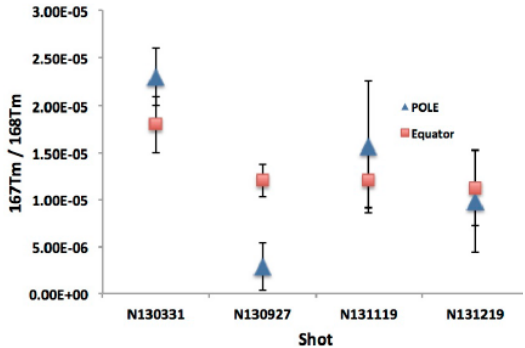}
\caption{The measured $^{167}$Tm/$^{168}$Tm ratios for the four shots which used the thicker (2 mm) foils.
The corresponding RIF($E_n>$15 MeV)/Total ratios are slightly shot dependent but are about 8 times larger than the Tm ratios. 
The factor of 8 comes from the difference in the shape of the 14 MeV and RIF spectra and the $(n,2n)$ and ($n,3n)$ cross sections for Tm.
 The shot dependence arises since the shape of the RIF spectrum depends on the density of the cold fuel. }
\end{figure}
The measured decays of the 208 keV signal in all of these experiments 
were consistent with the known $^{167}$Tm half life (Fig. 7).
\begin{figure}
\includegraphics[width=3in, bb= 0 0 720 540]{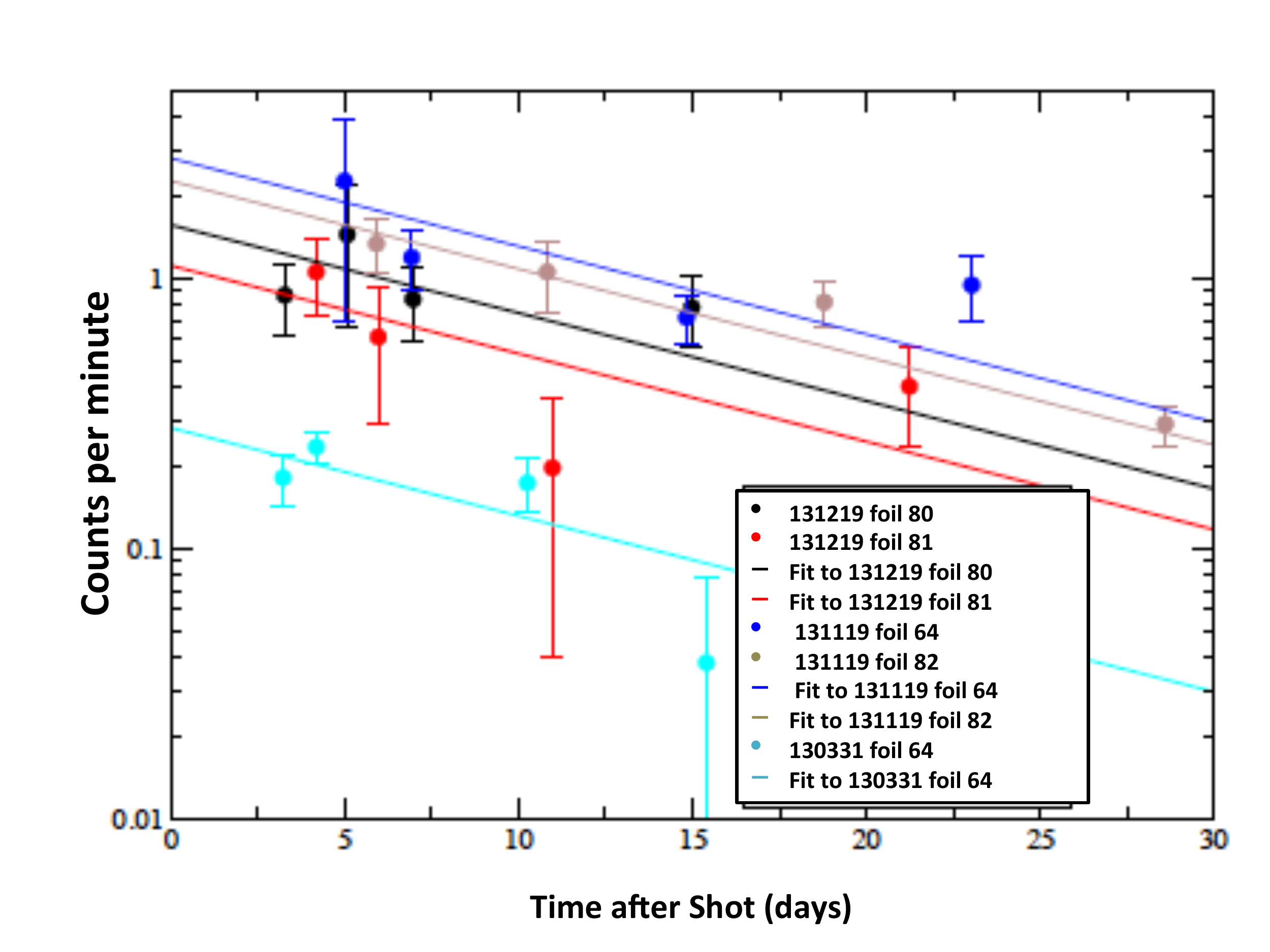}
\caption{The  208 keV signal for several shots as a function of the time since the shot. The decay rates of the signals are consistent with the 9.25 day half-life of $^{167}$Tm.}
\end{figure}

It is worth noting that for shot N130927 we saw a large asymmetry between the signal 
measured in the polar versus the equatorial directions, 
where the foil in the equatorial direction showed a $^{167}$Tm/$^{168}$Tm ratio of 1.25$\pm0.3\times10^{-5}$, 
while the polar signal was consistent with zero, $2.5\pm2.5\times 10^{-6}$.  The neutron image of the cold fuel for this same shot was also highly
asymmetric and it is not clear that the cold fuel was well assembled. 
This unexpectedly large asymmetry is still under theoretical investigation and we did not include this shot in our analysis of the stopping powers.
 
The approximate 35\% uncertainty in the RIF signal seen in the 2mm thick foils 
was largely due to self-attenuation of the 208 keV $\gamma$-ray signal within the foils, which limited the signal statistics.
For this reason we decided to move to thinner, 0.5 mm, foils. Typically, the thinner foils reduced the uncertainties to about 15\%.
Figure 8 shows the measured 208 keV peak for one such shot, shot N140304. 
Here the measured ratio of $^{167}$Tm/$^{168}$Tm was 1.69$\pm 0.24\times$10$^{-5}$. This lower uncertainty greatly reduces
the uncertainty in
extracting information on the stopping power in the cold fuel, so we will concentrate on this shot in our analysis below.
Again, the decay of the 208 keV signals seen in the 0.5 mm foils  is consistent with the half-life of $^{167}$Tm, Figure 9.
\begin{figure}
\includegraphics[width=3in, bb= 0 0 792 612]{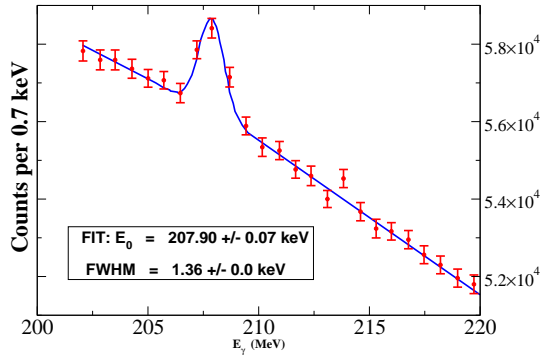}
\caption{The 208 keV peak from the decay of $^{167}$Tm for the shot N140304. The corresponding
measured ratio of $^{167}$Tm/$^{168}$Tm was 1.69$\pm 0.24\times$10$^{-5}$.}
\end{figure}
\begin{figure}
\includegraphics[width=3in, bb= 0 0 720 540]{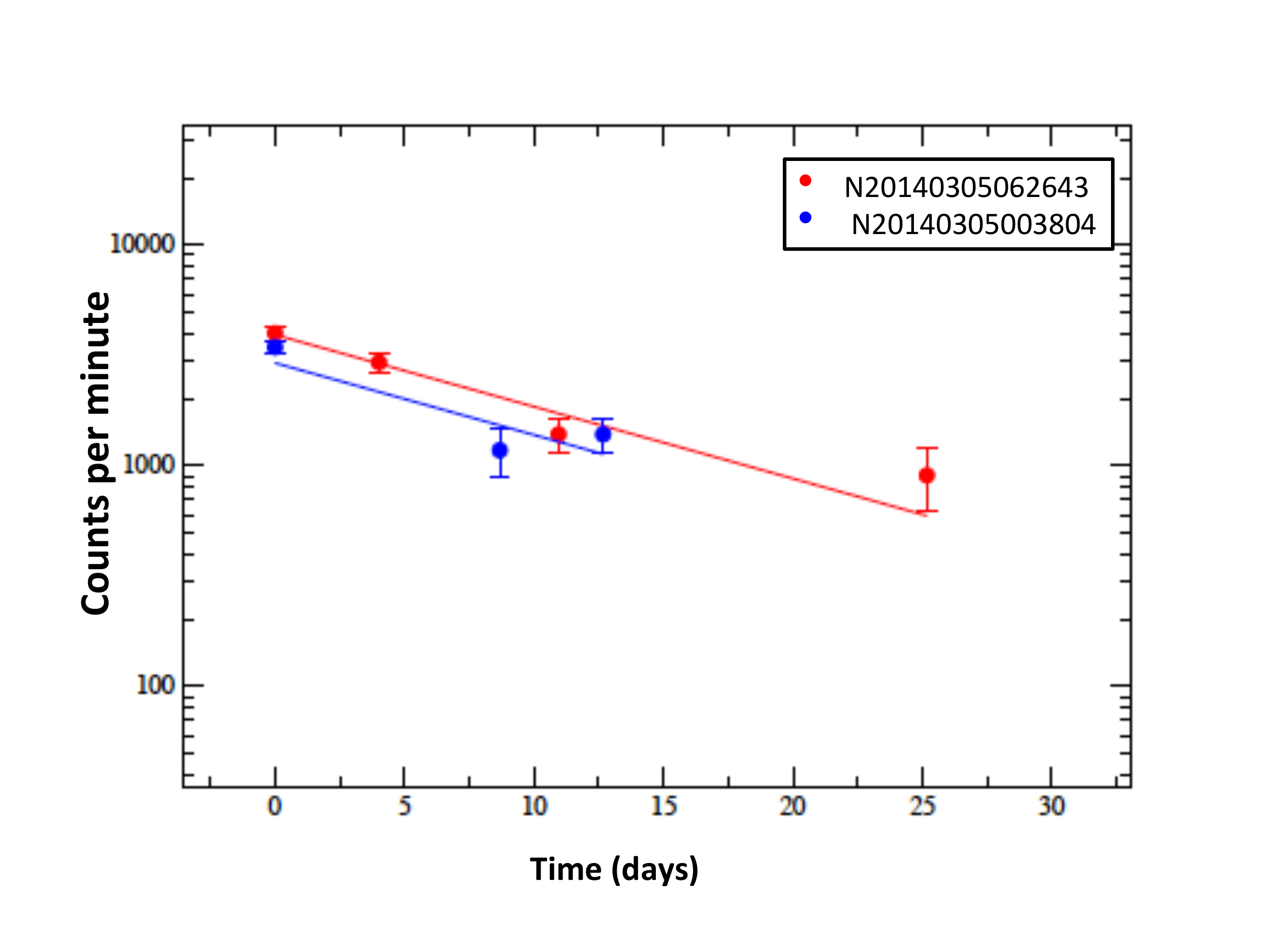}   
\caption{The decay of the 208 KeV signals from the foils for shots N140305, both consistent with the half-life of $^{167}$Tm.}
\end{figure}

\section{Extracting limits on the stopping power from the higher precision thin foil RIF measurements}
We first restrict our discussions to the cold fuel. We will add the contributions to the RIF production 
from the hotspot in the full analysis below.
Since the primary 14 MeV neutrons are only produced in the hot spot, the fluence of these neutrons at any position in the
cold fuel approximately follows a $1/(4\pi R^2)$ law. Thus, the volume and time integral of eq. 2 within the cold fuel can be written as
\begin{equation}
N^{CF}_{RIF}=N_{14}\left<\rho R\right>_{CF}\int_0^{E_{_{ko}}^{max}} \frac{d\tilde{\psi}_{ko}}{dE_{ko}}\sigma_{DT}\frac{dF}{dE_{RIF}}dE_{ko},
\label{tilde}
\end{equation}
where $\frac{d\tilde{\psi}_{ko}}{dE_{ko}}$ is the knock-on fluence per unit primary 14 MeV neutron, $N_{14}$ is the total 14 MeV yield, and $\left<\rho R\right>_{CF}$ is the areal density of the cold fuel.
In constraining stopping powers from the RIF measurements, our philosophy has been to use as much additional experimental information as possible. 
Thus, eq. \ref{tilde} has the advantage of displaying the explicit dependence on the yield and the areal density of the cold fuel.
The additional shot-dependent physics determining the RIF ratio is contained in variations of the density (and the related effective temperature) 
across the cold fuel, as well as possible 3-D cold fuel asymmetries.

In our analyses we used  full 3-D simulations constrained \cite{cerjan} by 
a broad range of  experimental data. The constraints involve 
a physically consistent description of the imploded capsule at stagnation. 
This method attempts to validate the model of the capsule through 
stringent comparisons between the radiation hydrodynamic code simulations and a suite of x-ray and neutron imaging data.
We carried out a second independent analysis of the RIFs 
using the 1-D code CPT-Implode, which also focuses on describing the stagnation properties of the capsule correctly. 
In both cases, we varied the stopping models and examined the change in the shape of the predicted RIF spectrum and its effect on
the
$^{167}$Tm/$^{168}$Tm ratio. 
We also calculated the RIF spectrum using eqs. 1 and 2, using the density and
temperature profiles from the 1-D simulation. The three methods agreed to about 20\%, 
providing confidence in the formalism
and the charge-particle transport packages in the codes.

In Table 1, we list the measured and predicted values for the $^{167}$Tm/$^{168}$Tm diagnostic for two shots, comparing different stopping model predictions. 
As can be seen, the measurements do not agree with either the Corman-Spitzer \cite{corman,spitzer} or Grabowski \cite{paul} models.
The higher accuracy measurements, which were obtained using the thinner (0.5 mm) foils, favor the Maynard-Deutsch model over the Li-Petrasso model. 
The BPS model, which is only designed for weakly coupled plasma, predicts a stopping power with an unphysical sign.
\begin{table}
\caption{Comparison between the measured and predicted RIF signals for two shots.}
\begin{tabular}{c|c}\hline
1$^{st}$ RIF observation&\\
Shot N130331   & $^{167/168}$Tm\\\hline
Measurement& 2.03$\pm$0.75$\times$10$^{-5}$\\
(averaged over both foils)&\\
Maynard-Deutsch \cite{MD}& 1.99$\times$10$^{-5}$\\
Li-Petrasso \cite{LiP}& 1.57$\times$10$^{-5}$\\
Corman-Spitzer \cite{corman, spitzer}& 0.55$\times$10$^{-5}$\\\hline
&\\
Most accurate measurement to-date&\\
Shot N140304&                  $^{167/168}$Tm\\\hline
Measurement& 1.69$\pm$0.24$\times$ 10$^{-5}$\\
(only one foil was used in this shot)&\\
Maynard-Deutsch \cite{MD}& 1.59$\times$10$^{-5}$\\
Li-Petrasso \cite{LiP}& 1.16$\times$10$^{-5}$\\
Grabowski \cite{paul}&        0.89$\times$10$^{-5}$\\
BPS \cite{BPS}& Model breaks-down\\ 
&$(\Gamma$ too large)\\\hline
\end{tabular}
\end{table}

We found that the effect of electron degeneracy on the predicted RIF signal depended on the stopping model. 
Degeneracy is most important
at low knock-on energies, where the stopping power scales with $n_e/\theta_\mathrm{eff}^{3/2}$. 
The inclusion of 
electron degeneracy increased the Tm diagnostic anywhere from a factor of 1.5-3.0, depending on the stopping model. 
A more precise measure of the effect of degeneracy would be afforded by a direct measurement of the shape of the RIF spectrum by neutron time-of-flight.  
We expect the cold fuel degeneracy to cause a distortion of the spectrum, with an enhancement at lower energies, relative to the predictions  with no treatment of degeneracy.

Finally, as discussed above, a small fraction  of the RIFs were predicted to originate in the hotspot. 
The hotspot plasma is a weakly coupled plasma for which the form of the stopping power is known. The hotspot has an additional contribution to RIF production from alpha-particle induced knock-on ions.
Our simulations predict that the combination of the neutron-induced and alpha-induced RIFS from the hotspot make up about 10-15\% of the measured $^{167}$Tm/$^{168}$Tm signal. 
These RIF contributions are included in Table 1.

\section{summary}
The  measurements presented here represent the first observation of RIF neutrons in any inertial confinement fusion system.
RIFs were observed in both Low Foot and High Foot design cryogenic capsules, with a magnitude of about $10^{-4}$ of the total neutron spectrum. 
The majority of the RIF neutrons were produced in the
dense cold fuel surrounding the burning hotspot of the capsule, 
which is moderately to strongly coupled $(\Gamma\sim$0.3-1.2) and electron degenerate $(\theta_\mathrm{Fermi}/\theta_e\sim$2.3-5.0).
Thus, the RIF data presented here provide singular tests of stopping power under warm dense plasma conditions.
Future direct measurements of RIF spectra produced in cryogenic NIF capsules 
by time-of-flight would place even more stringent constraints on energy loss in strongly coupled systems, particularly
with respect to the effects of plasma degeneracy.


\begin{thebibliography}{99}
\bibitem{NIF}M. J. Edwards {\it et al.}, Phys. Plasmas {\bf 20}, 070501 (2013).
\bibitem{MFE}R.K. Fisher,P.B. Park, J.M. McChesney, and M.N. Rosenbluth, Nuclear Fusion {\bf 34}, 1291 (1994), and\\
J. K\"allne, L. Ballabio, J. Frenje, S. Conroy, G. Ericsson, M. Tardocchi, E. Traneus, and G. Gorini, Phys. Rev. Letts. {\bf 85}, 1246(2000), and\\
L. Ballabio, G. Gorini, and J. K\"allne,  Phys. Rev. E {\bf 55} , 3358 (1997), and\\
A. A. Korotkov, A. Gondhalekar, and R. J. Akers, Phys. of Plasmas 7, 957 (2000). 
\bibitem{stopping-jungman}The Relationship between Charged-Particle Fluence and Stopping Power, Gerard Jungman and A.C.Hayes, Los Alamos Internal Report, LA-UR-13-26171, (2013).
\bibitem{MD} G.Maynard and C. Deutsch, J. Physique, 46, 1113 (1985).
\bibitem{zim}G.B. Zimmerman, Recent Developments in Monte Carlo Techniques, Lawrence Livermore National Laboratory internal report, UCRL-JC-105616, (1990).
\bibitem{paul} Paul E. Grabowski, Michael P. Surh, David F. Richards, Frank R. Graziani, and Michael S. Murillo, Phys. Rev. Lett. 111, 215002, (2013).
\bibitem{beta-mix}A.C. Hayes, Gerard Jungman, J.C. Solem, P.A. Bradley, and R.S. Rundberg,
Modern Physics Letters A {\bf 21}, No. 13 (2006) 1029.
\bibitem{hayes-rif} A.C. Hayes and Gerard Jungman, The Relationship between Knock-on Charged-Particle Reactions and
Reaction-in-Flight Neutrons, Los Alamos Internal Report, LA-UR-13-27001, (2013).
\bibitem{density}For the sake of brevity, we use the symbol $n_{DT}$ to mean both the deuteron and triton density. IF the knock-on ion is a deuteron, $n_{DT}$ represents the deuteron density in the expression for $Q_0$, and the density of the triton in eq. \ref{bigeqn}, and vise-verse.
\bibitem{cpt-burn} CPT-Implode is a 1-D ICF  burn code based on the burn physics of \cite{French}. CPT refers to
charged-particle transport as the code places and emphasis on knock-on ions and the production of RIFs.
\bibitem{hydra-burn}M. M. Marinak, G. D. Kerbel, N. A. Gentile, O. Jones, D. Munro, S.Pollaine,
T. R. Dittrich, and S. W. Haan,
Phys. Plasmas {\bf 8}, 2275 (2001)
\bibitem{BPS} L. S. Brown, D. L. Preston, and R. L. Singleton, Jr., Phys. Rep. 410, 237 (2005).
\bibitem{Zwick} G\"unter Zwicknagel, Theory and simulation of heavy ion stoping in plasma, Laser and Particle Beams {\bf 27} (2009) 399-413.
\bibitem{stopping-test} A.C. Hayes, Accuracy of the BPS and Maynard-Deutsch Stopping Powers, Internal Los Alamos Report, LA-UR-13-22639, (2013).
\bibitem{highfoot}T. R. Dittrich, O. A. Hurricane, D. A. Callahan, E. L. Dewald, T. Döppner, D. E. Hinkel, L. F. Berzak Hopkins, S. Le Pape, T. Ma, J. L. Milovich, J. C. Moreno, P. K. Patel, H.-S. Park, B. A. Remington, J. D. Salmonson, and J. L. Kline,
Phys. Rev. {\bf 112}, 055002 (2014), and\\
H.-S. Park, O. A. Hurricane, D. A. Callahan, D. T. Casey, E. L. Dewald, T. R. Dittrich, T. Döppner, D. E. Hinkel, L. F. Berzak Hopkins, S. Le Pape, T. Ma, P. K. Patel, B. A. Remington, H. F. Robey, J. D. Salmonson, and J. L. Kline, Phys. Rev. Lett. {\bf 112}, 055001 (2014), and\\
    O. A. Hurricane, {et al.}, Phys. Plasmas, {\bf 21}, 056314 (2014).
\bibitem{LiP}C. K. Li and R. D. Petrasso, Phys. Rev. Lett. 70, 3059 (1993)
\bibitem{Hayes-pop}A. C. Hayes, P. A. Bradley, G. P. Grim, Gerard Jungman, and J. B. Wilhelmy,PHYSICS OF PLASMAS 17, 012705,2010.
\bibitem{Dawn} D.A. Shaughnessy, {\it et al.}, Rev. Sci. Instrum. {\bf 83} , 10D917 (2012).
\bibitem{gary} G.P. Grim, {\it et al.}, SPIE conference proceedings, {\it in press}.
\bibitem{cerjan} Charles Cerjan, Paul T. Springer, and Scott M. Sepke, Phys. Plasmas 20, 056319 (2013).
\bibitem{corman} E.G. Corman, W.E. Loewe, G.E. Cooper, and A.M. Winslow, Nucl. Fus, {\bf 15} 377 (1975).
\bibitem{spitzer} L. Spitzer, {\it The Physics of Fully Ionized Gases}, New York: Interscience, (1965).
\bibitem{French}J. Sanz, J. Garnier, C. Cherfils, B. Canaud, and L. Masse, M. Temporal, Phy. Plasmas {\bf 12}, 11272005.
\end{thebibliography}
\end{document}